# Successive Cancellation List Polar Decoder using Log-likelihood Ratios

Bo Yuan and Keshab K. Parhi, *Fellow*, *IEEE*

Department of Electrical and Computer Engineering, University of Minnesota Twin Cities

***Abstract*—Successive cancellation list (SCL) decoding algorithm is a powerful method that can help polar codes achieve excellent error-correcting performance. However, the current SCL algorithm and decoders are based on likelihood or log-likelihood forms, which render high hardware complexity. In this paper, we propose a log-likelihood-ratio (LLR)-based SCL (LLR-SCL) decoding algorithm, which only needs half the computation and storage complexity than the conventional one. Then, based on the proposed algorithm, we develop low-complexity VLSI architectures for LLR-SCL decoders. Analysis results show that the proposed LLR-SCL decoder achieves 50% reduction in hardware and 98% improvement in hardware efficiency.**

***Keywords*—polar codes, successive cancellation list (SCL), log-likelihood ratio (LLR), VLSI, low-complexity**

## I. INTRODUCTION

Polar codes have become one of the most favorable forward error correction (FEC) codes since their discovery in 2008 [1]. Nowadays information theorists have shown that the capacity-achieving polar codes can achieve beyond-LDPC error-correcting performance with the use of successive cancellation list (SCL) algorithm [2]. Therefore, the SCL algorithm is viewed as the most promising approach for practical polar decoding.

To date, original SCL algorithm and its variants are based on likelihood [2] or log-likelihood (LL) forms [3-4][8]. Because most FEC-contained systems are based on log-likelihood ratio (LLR) form, the current non-LLR-based SCL decoders are incompatible for practical applications. Moreover, because two types of decoding messages (probability being 0 and 1) need to be processed and stored in non-LLR-based SCL decoders, the corresponding hardware complexity is very high.

This paper presents an LLR-based SCL (LLR-SCL) algorithm for polar code decoding. In the proposed algorithm, the ratio of probability being 0 and 1 is used to represent the decoding messages. As a result, the computation complexity and memory requirement of LLR-SCL algorithm is greatly reduced as compared to LL-SCL case. Then, based on this new algorithm, a VLSI architecture of the LLR-SCL decoder is presented. Analysis shows that the proposed LLR-SCL decoder achieves 50% reduction in hardware and 98% increase in hardware efficiency.

The rest of this paper is organized as follows. Section II reviews the polar codes. The proposed LLR-SCL algorithm is presented in Section III. Section IV develops the hardware architecture of LLR-SCL decoder. Hardware performance is analyzed in Section V. Section VI draws the conclusions.

## II. REVIEW OF POLAR CODES

### *A. Polar Code*

As shown in [1], the reliability of decoded bits over discrete binary memoryless symmetric channel (D-BMS) can be polarized according to their positions at the codeword. Therefore, by assigning $k$ bits in the source data and ($n$-$k$) "0" bits over the reliable and unreliable positions, respectively, we can construct a length-$n$ polar code with rate $R$=$k$/$n$. In general, these ($n$-$k$) "0" bits are called "frozen" bits while the $k$ information bits are called "free" bits. For the details of polar encoding, the reader is referred to [1].

### *B. Successive Cancellation SC Algorithm*

An SCL decoder can be viewed as multiple copies of successive cancellation (SC) decoder; therefore we first introduce SC decoder in this subsection.

Fig. 1 shows the decoding scheme of a likelihood-based SC decoder with $n$=4. Here the SC decoder consists of $\log_2 n$=2 stages, where each stage consist of 4-input 2-output **f** and **g** units as the basic computation units. At the end of the last stage (stage2 in this example), a hard-decision unit (**h**) is used to determine the decoded bit $\hat{u}_i$. Notice that in Fig. 1 each **f** or **g** unit is labeled with a number, which indicates the time index when the corresponding unit is activated.

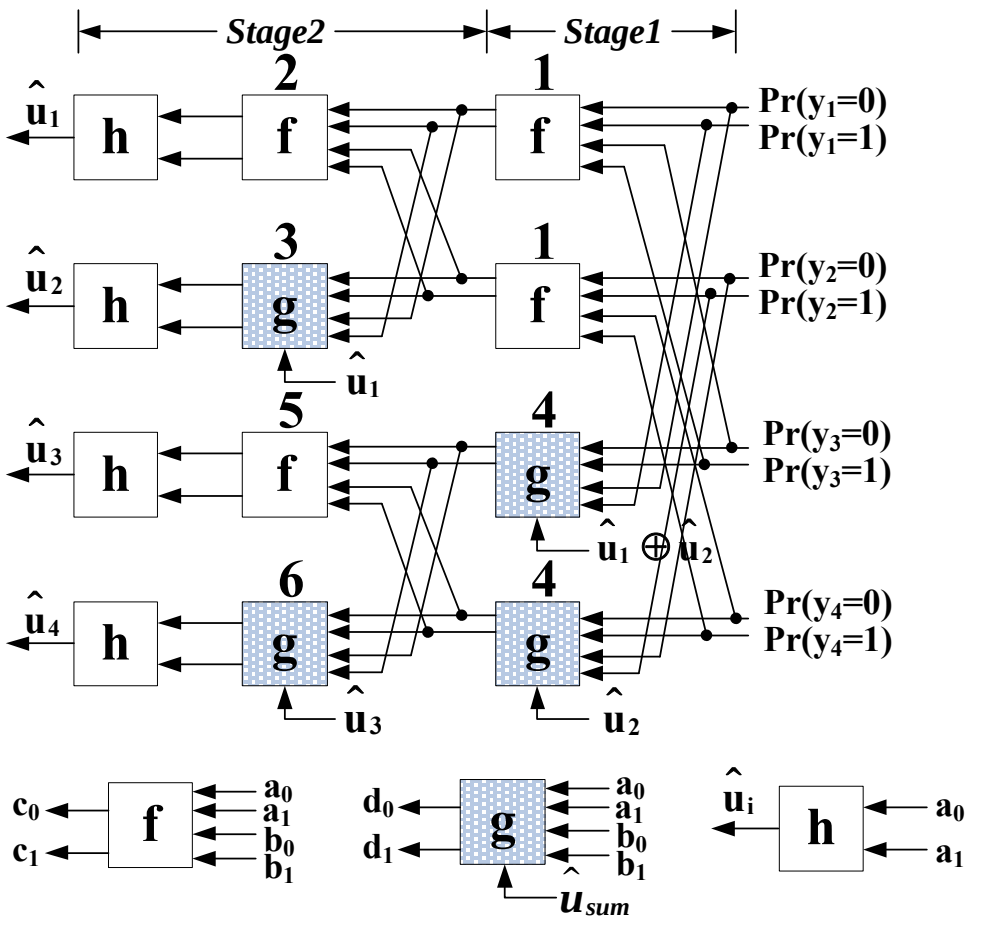

Fig. 1. Decoding scheme of $n$=4 likelihood-based SC decoder, cited from [8].

The function of **f** and **g** units can be derived from the polar encoding procedure. Fig. 2(a) shows the basic computation unit of the polar encoder. It can be seen that the basic encoding computation is a left-to-right transformation as $out_1$=$in_1 \oplus in_2$, and $out_2$=$in_2$, where $\oplus$ is the exclusive-or operation.

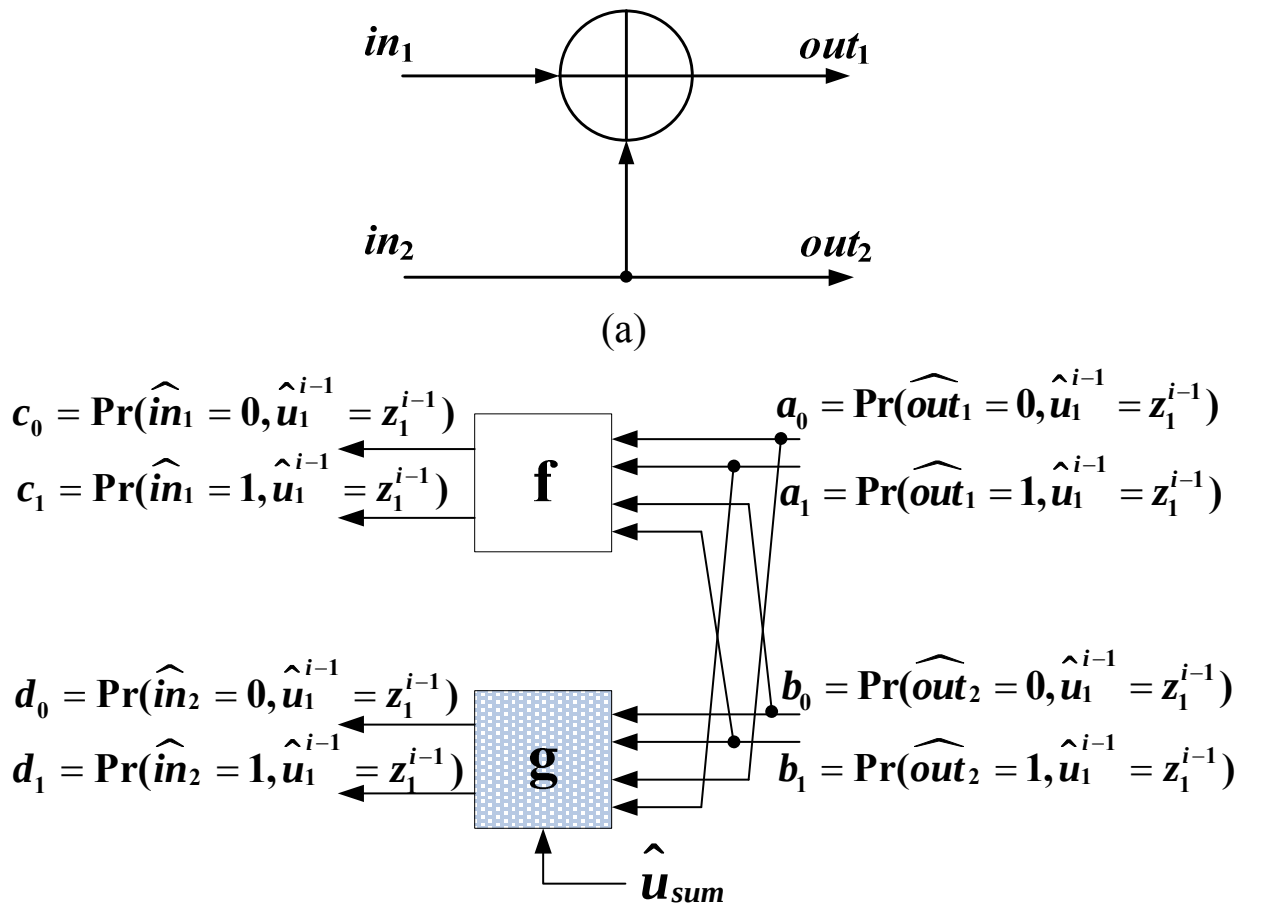

Fig. 2. (a) basic unit of polar encoding. (b) basic unit of polar SC decoding, cited from [8].

Correspondingly, the functions of **f** and **g** units, as the basic computation unit of polar decoder, represent right-to-left estimations from $\widehat{out}_1$ and $\widehat{out}_2$ to $\widehat{in}_1$ and $\widehat{in}_2$, where $\widehat{in}_1$, $\widehat{in}_2$, $\widehat{out}_1$, $\widehat{out}_2$ denote the estimates of $in_1$, $in_2$, $out_1$ and $out_2$, respectively. If we assume previous decoded bits $\hat{u}_1$, $\hat{u}_2$ … $\hat{u}_{i-1}$ are $z_1$, $z_2$ … $z_{i-1}$, and denote this event as $\hat{u}_1^{i-1} = z_1^{i-1}$, then we can derive the functions of **f** and **g** units as follows:

$$c_0 = \Pr(\widehat{in}_1 = 0, \hat{u}_1^{i-1} = z_1^{i-1}) = a_0 b_0 + a_1 b_1 \tag{1}$$

$$c_1 = \Pr(\widehat{in}_1 = 1, \hat{u}_1^{i-1} = z_1^{i-1}) = a_0 b_1 + a_1 b_0 \tag{2}$$

$$d_0 = \Pr(\widehat{in}_2 = 0, \widehat{in}_1 = \hat{u}_{sum}, \hat{u}_1^{i-1} = z_1^{i-1}) = a_{\hat{u}_{sum}} b_0 \tag{3}$$

$$d_1 = \Pr(\widehat{in}_2 = 1, \widehat{in}_1 = \hat{u}_{sum}, \hat{u}_1^{i-1} = z_1^{i-1}) = a_{1-\hat{u}_{sum}} b_1 , \tag{4}$$

where $a_0 = \Pr(\widehat{out}_1 = 0, \hat{u}_1^{i-1} = z_1^{i-1})$, $a_1 = \Pr(\widehat{out}_1 = 1, \hat{u}_1^{i-1} = z_1^{i-1})$, $b_0 = \Pr(\widehat{out}_2 = 0, \hat{u}_1^{i-1} = z_1^{i-1})$, and $b_1 = \Pr(\widehat{out}_2 = 1, \hat{u}_1^{i-1} = z_1^{i-1})$ are the 4 inputs of **f** and **g** units.

Besides, the function of hard-decision **h** unit is

$$\hat{u}_i = \begin{cases} 0 & \text{if } a_0 \ge a_1 \text{ or } \hat{u}_i \text{ is frozen bit} \\ 1 & \text{if } a_0 < a_1 \text{ and } \hat{u}_i \text{ is free bit} \end{cases} \tag{5}$$

In general, (1)-(5) describe likelihood-based SC algorithm.

### *C. SC Algorithm over code tree*

On the other hand, from the view of code tree, the SC algorithm can be viewed as the path searching process over $n$-level code tree. Fig. 3 shows an example searching procedure with $n$=4. Here level-$i$ represents decoded bit $\hat{u}_i$. In addition, the value associated with each node is the metric for the path from root node to the current node. For example, 0.09 is the path metric for length-3 path (1,0,1). Here (1,0,1) represents $\hat{u}_1 = 1$, $\hat{u}_2 = 0$ and $\hat{u}_3 = 1$. Therefore, the path (1,0,1) with metric 0.09 indicates that $\Pr(\hat{u}_1 = 1, \hat{u}_2 = 0, \hat{u}_3 = 1) = 0.09$.

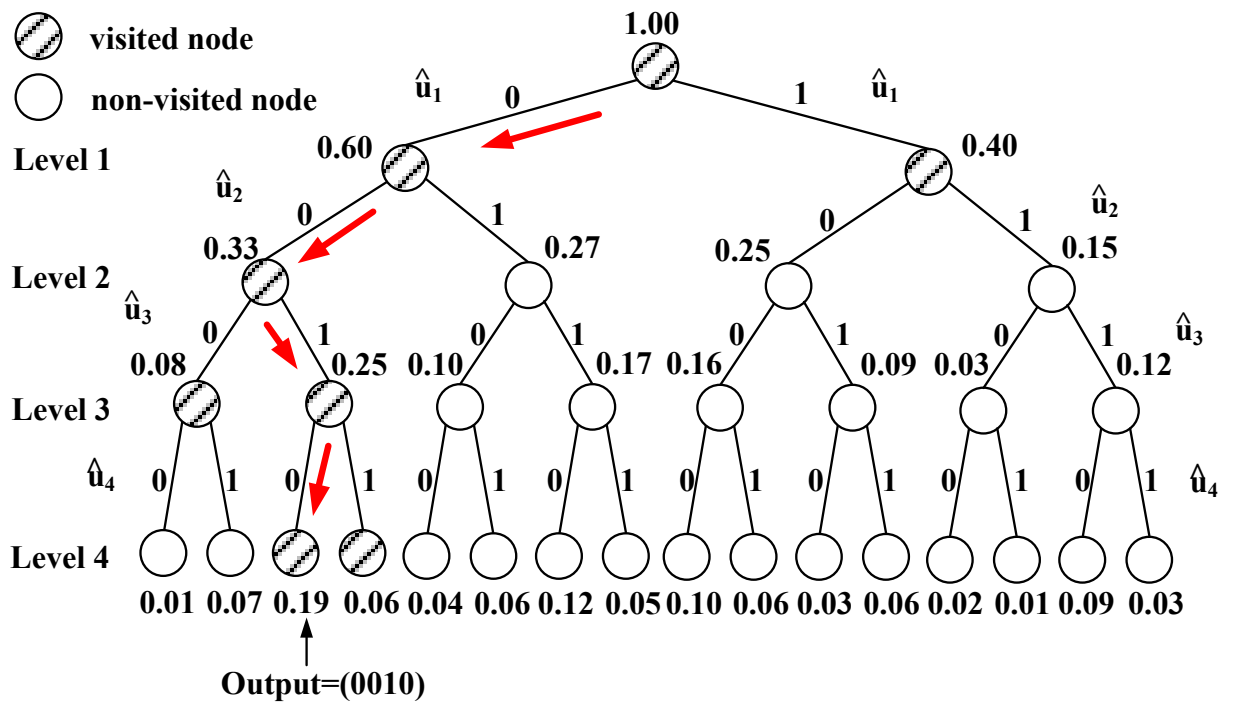

Fig. 3. Example SC searching process over code tree for $n$=4 polar code, cited from [8].

Notice that for the **f** or **g** units in the last stage (for example stage2 in Fig. 1), $\widehat{in}_1$ or $\widehat{in}_2$ is $\hat{u}_i$. In that case, according to (1)-(5), $c_0$, $c_1$, $d_0$ or $d_1$, as the output of **f** or **g** unit, represent the joint probabilities of $\hat{u}_1$, $\hat{u}_2$ …, $\hat{u}_i$, which is just the metric of path ( $\hat{u}_1$, $\hat{u}_2$ …, $\hat{u}_i$ ). Therefore, the path metric of SC algorithm is calculated by the **f** or **g** units in the last stage.

With the use of path metric, the SC decoder performs a locally optimal searching strategy to find the length-$n$ path with largest metric. As shown in Fig. 3, in each level the SC decoder first visits two children nodes that are associated with the current survival path. By comparing the corresponding path metrics, the SC decoder selects the path with larger metrics as the updated survival path. The example survival path in Fig. 3 is marked as red arrows. It can be seen that the valid length-4 path with largest metric 0.19 can be found by the SC decoder.

### *D. SCL Decoding Algorithm*

Due to the limit of locally optimal search, in many cases the SC decoder cannot find the correct decoding path. In [2] SCL algorithm was proposed to solve this problem. By using multiple SC decoders over the same code tree, the chance of finding the correct decoding path is significantly improved. Here the number of SC component decoders is referred as list size $L$. Fig. 4 shows an example of $n$=4 SCL decoder with $L$=2. It can be seen that the SCL decoder can trace the correct path (1,0,0,0) with metric 0.23 while SC decoder cannot.

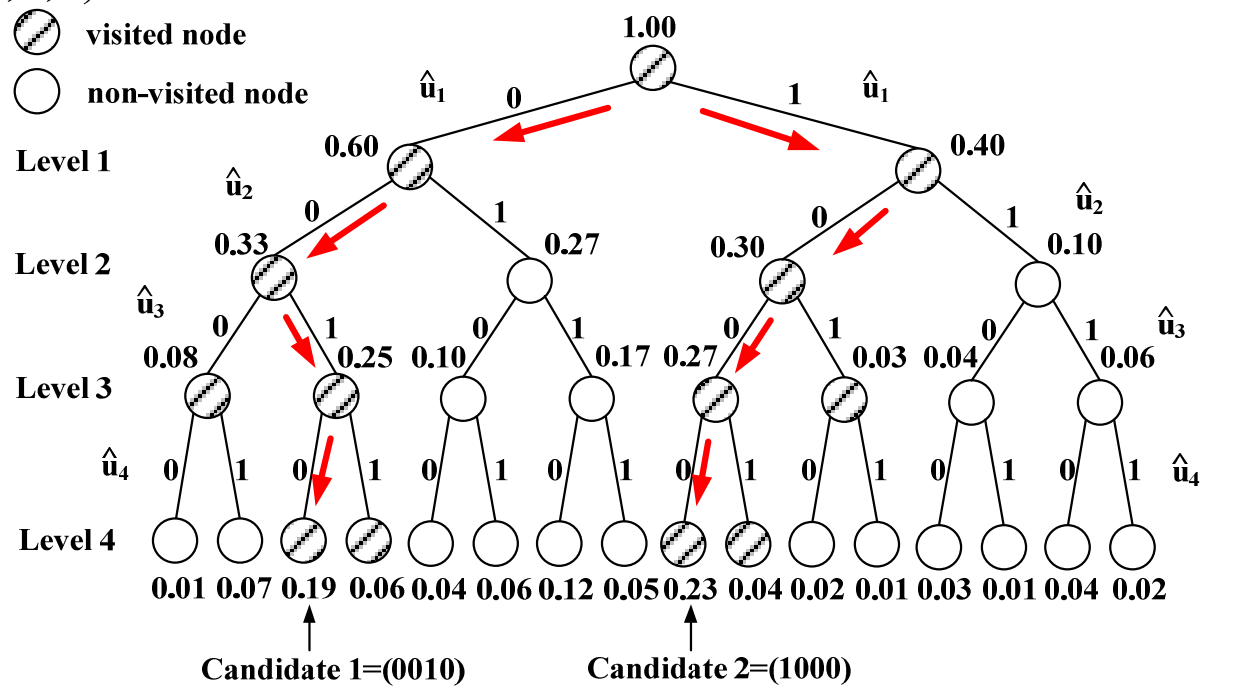

Fig. 4. Example SCL searching process over code tree for $n$=4 polar code with $L$=2, cited from [8].

## III. THE PROPOSED LLR-SCL ALGORITHM

### A. Benefit of LLR-based Representation

In Section II we present the likelihood-based SC and SCL algorithms. However, in practical applications LLR-based representation, instead of likelihood-based form, is usually adopted for soft FEC decoder designs. This is because the LLR-based designs have much less hardware complexity than the non-LLR-based ones. Generally, in order to describe the joint probabilistic information of a bit $v$ and event $\Psi$, the likelihood-based decoders need to process and store two types of messages as Pr($v$=0, $\Psi$) and Pr($v$=1, $\Psi$), while LLR-based decoders only need to deal with one type of message as (ln(Pr($v$=0, $\Psi$)/Pr($v$=1, $\Psi$))). As a result, the computation complexity and memory requirement of LLR-based decoders are much lower than their likelihood-based counterparts.

In [5], the LLR-based representation had been used in SC decoder design. The success of LLR-based scheme in SC algorithm is built on the property that the binary value of decoded bit $\hat{u}_i$ can be directly determined from the sign of corresponding LLR messages. However, in SCL algorithm no such property exists. The $\hat{u}_i$ in the SCL algorithm has to be determined by comparing the metrics of all the candidate paths, which are inherently based on likelihood form. As a result, the current SCL decoders are either based on likelihood or log-likelihood (LL) form, instead of LLR form.

### B. The Proposed LLR-SCL Decoding Algorithm

In this subsection we present a LLR-based SCL (LLR-SCL) algorithm. First, we convert original likelihood-base messages in (1)-(4) to the LLR-based forms as follows:

$$c = 2\tanh^{-1}(\tanh(\frac{a}{2})\tanh(\frac{b}{2})) \approx sign(a)sign(b)\min(|a|,|b|) \quad (6)$$

$$d = a(-1)^{\hat{u}_{sum}} + b \quad (7)$$

where $c$=ln($c_0/c_1$), $d$=ln($d_0/d_1$), $a$=ln($a_0/a_1$) and $b$=ln($b_0/b_1$).

After representing all the messages in the LLR form, the next step is to calculate path metrics, which is the key task for developing LLR-based SCL algorithm. For the likelihood-based SCL algorithm, this calculation is automatically performed by the likelihood-based **f** or **g** unit in the last stage of SC component decoders. However, for the LLR-SCL algorithm, after the LLR-based **f** or **g** unit outputs LLR messages $c$ or $d$, an extra metric computation unit (MCU) is needed to calculate path metrics. Next we derive the function of MCU.

First, notice that the metric for the length-$i$ path ($\hat{u}_1$, $\hat{u}_2$ …, $\hat{u}_i$) is $\Pr(\hat{u}_1 = z_1, \ldots \hat{u}_{i-1} = z_{i-1}, \hat{u}_i = z_i) = \Pr(\hat{u}_i = z_i, \hat{u}_1^{i-1} = z_1^{i-1})$. Then with the log-domain representation, the metrics for length-$i$ paths are:

$$M_{i,0} = \ln(\Pr(\hat{u}_i = 0, \hat{u}_1^{i-1} = z_1^{i-1})) \quad (8)$$

$$M_{i,1} = \ln(\Pr(\hat{u}_i = 1, \hat{u}_1^{i-1} = z_1^{i-1})), \quad (9)$$

where $M_{i,0}$ and $M_{i,1}$ represent the path metrics when $\hat{u}_i$ = 0 and 1, respectively.

Then, recall the function of MCU is to calculate the path metrics $M_{i,0}$ and $M_{i,1}$ with the known $c$ or $d$ output from **f** or **g** unit in the last stage. Without loss of generality, we assume MCU uses $c$ to calculate path metrics. As a result, for the **f** units in the last stage, we have:

$$c = \ln\frac{c_0}{c_1} = \ln\frac{\Pr(\hat{u}_i = 0, \hat{u}_1^{i-1} = z_1^{i-1})}{\Pr(\hat{u}_i = 1, \hat{u}_1^{i-1} = z_1^{i-1})} = M_{i,0} - M_{i,1} \quad (10)$$

In addition, similar to the case for length-$i$ path, the log-domain metric $M_{i-1,z_{i-1}}$ for length-($i$-1) path ( $\hat{u}_1 = z_1$ , $\hat{u}_2 = z_2$ …, $\hat{u}_{i-1} = z_{i-1}$ ) can be represented as:

$$\begin{aligned} M_{i-1,z_{i-1}} &= \ln(\Pr(\hat{u}_{i-1} = z_{i-1}, \hat{u}_1^{i-2} = z_1^{i-2})) = \ln(\Pr(\hat{u}_1^{i-1} = z_1^{i-1})) \\ &= \ln(\Pr(\hat{u}_i = 0, \hat{u}_1^{i-1} = z_1^{i-1}) + \Pr(\hat{u}_i = 1, \hat{u}_1^{i-1} = z_1^{i-1})) \\ &= \ln(e^{M_{i,0}} + e^{M_{i,1}}) \end{aligned} \quad (11)$$

Consequently, based on (10)(11) we can calculate $M_{i,0}$ and $M_{i,1}$ as follows:

$$M_{i,0} = M_{i-1,z_{i-1}} + c - \ln(1+e^c) \quad (12)$$

$$M_{i,1} = M_{i-1,z_{i-1}} - \ln(1+e^c) \quad (13)$$

The above (12)(13) contains **ln(•)** computation, which needs complex lookup table (LUT) for hardware design. Hence we need to simplify the calculations of $M_{i,0}$ and $M_{i,1}$.

Consider $\ln(1+e^x) \approx \begin{cases} x & \text{for large } x \\ 0 & \text{for small } x \end{cases}$, (12)(13) can be further approximated as follows:

$$M_{i,0} \approx \begin{cases} M_{i-1,z_{i-1}} & \text{if } c \geq 0 \\ M_{i-1,z_{i-1}} + c & \text{if } c < 0 \end{cases} \quad (14)$$

$$M_{i,1} \approx \begin{cases} M_{i-1,z_{i-1}} - c & \text{if } c \geq 0 \\ M_{i-1,z_{i-1}} & \text{if } c < 0 \end{cases} \quad (15)$$

Similarly, if MCU uses LLR messages $d$ to calculate path metrics, then we have:

$$M_{i,0} \approx \begin{cases} M_{i-1,z_{i-1}} & \text{if } d \geq 0 \\ M_{i-1,z_{i-1}} + d & \text{if } d < 0 \end{cases} \quad (16)$$

$$M_{i,1} \approx \begin{cases} M_{i-1,z_{i-1}} - d & \text{if } d \geq 0 \\ M_{i-1,z_{i-1}} & \text{if } d < 0 \end{cases} \quad (17)$$

In general, (14)-(17) show how to calculate the metric of length-$i$ path with LLR messages. After **f** or **g** unit in the last stage outputs the LLR message ($c$ or $d$), with the knowledge of length-($i$-1) path metric ( $M_{i-1,z_{i-1}}$ ), we can calculate the metric of length-$i$ path ($M_{i,0}$ and $M_{i,1}$). As a result, the LLR-SCL algorithm is summarized as follows:

**Scheme A: LLR-based (n, k) SC list decoding (LLR-SCL) with list size L**

1: **Input:** *Log - Likihood ratios of each bit in the received codeword*
2: **Initilization:** *Path metric* $M_0 = 0$ *for each survival path*
3: **For** $i = 1$ **to** $n$
4:   **For each length-**$(i-1)$ **survival path** $(\hat{u}_1, ...\hat{u}_{i-1}) \triangleq z_1^{i-1}$ **with metric** $M_{i-1,z_{i-1}}$
5:     **SC decoding:** *Apply LLR - based SC to calculate LLRs output from last stage*
6:     **Metric Computation & Path Expansion:**
7:     *Expand survival path* $z_1^{i-1}$ *to* 2 *candidate paths* $(z_1^{i-1}, \hat{u}_i = 0)$ *and* $(z_1^{i-1}, \hat{u}_i = 1)$
8:     *1 length -* $(i-1)$ *path* $\Rightarrow$ 2 *length -* $i$ *paths*
9:     *LLRs output from last stage* **f** *(or* **g** *unit) is* $c$ *(or* $d$*)*
10:       **if** $c$ (or $d$) $\geq 0$   *path* $(z_1^{i-1}, 0)$ *with metric* $M_{i,0} = M_{i-1,z_{i-1}}$
11:           *path* $(z_1^{i-1}, 1)$ *with metric* $M_{i,1} = M_{i-1,z_{i-1}} - c$ *(or* $-d$*)*
12:       **else**     *path* $(z_1^{i-1}, 0)$ *with metric* $M_{i,0} = M_{i-1,z_{i-1}} + c$ *(or* $+d$*)*
13:           *path* $(z_1^{i-1}, 1)$ *with metric* $M_{i,1} = M_{i-1,z_{i-1}}$.
14:   **End for**
15:   **If** $\hat{u}_i$ **is free bit**
16:     **Compare and Prune:**
17:     *Compare the metrics of all the* $2L$ *length - (i) candidate paths*
18:     *Select* $L$ *paths with the* $L$ *largest metrics as the new survival paths*
19:   **else** ($\hat{u}_i$ **is frozen bit "0"**)
20:     *Keep all the path* $(z_1^{i-1}, 0)$ *as the new survival paths.*
21: **End for**
22: **Output:** *Choose the length - n survival path with the largest metric*

### *C. Simulation Results*

In subsection III-B, we perform approximation on the path metric calculation to avoid complex **ln(•)** computation. Fig. 5 shows this approximation does not cause performance loss. In addition, it is also seen that the approximated LLR-SCL algorithm has the same error-correcting performance with the original non-LLR-based SCL algorithm.

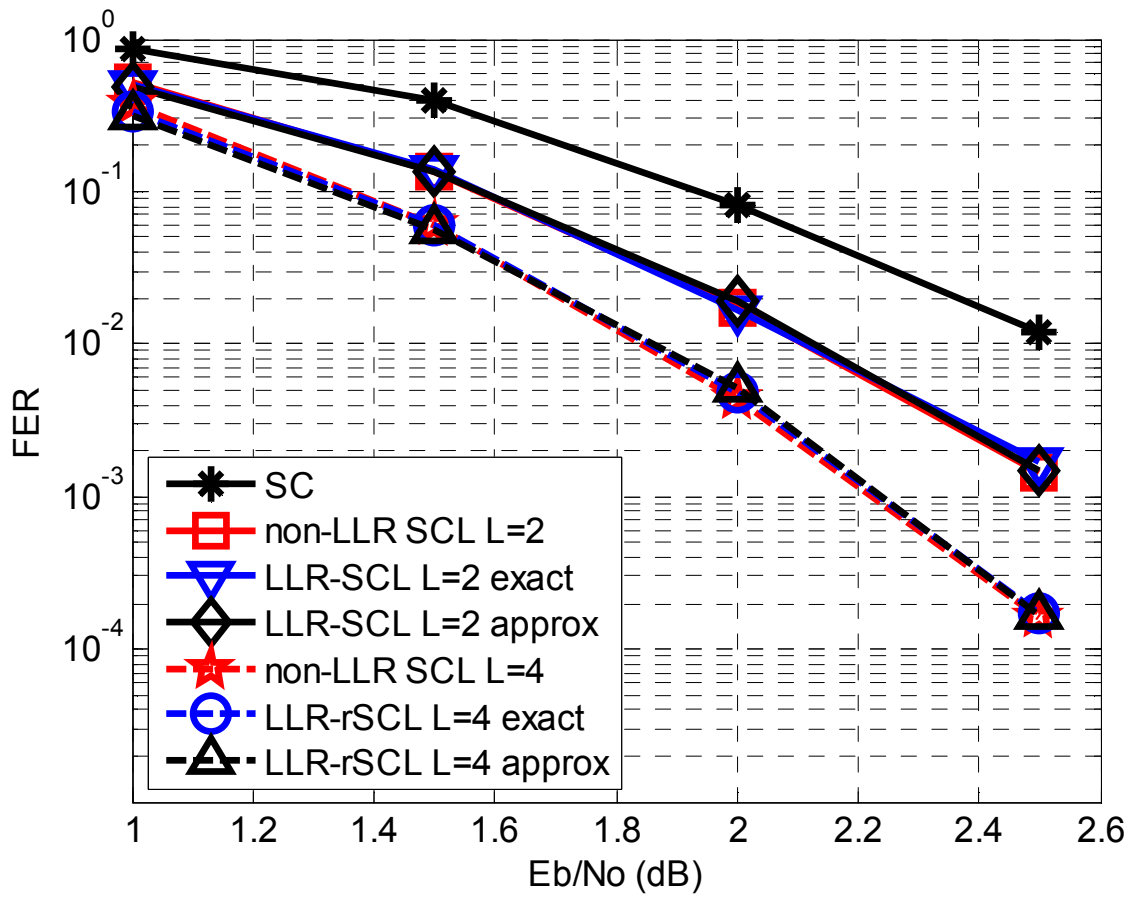

Fig. 5. Simulation results for polar (1024, 512) codes over AWGN channel.

## IV. Hardware Architecture of LLR-SCL Decoder

### *A. Overall Architecture*

In this section, based on the new LLR-SCL algorithm, we develop the corresponding hardware architecture. Fig. 6 shows the overall architecture of $L$-size LLR-SCL decoder. It can be seen that the LLR-SCL decoder consists of $L$ LLR-SC component decoders, which had been discussed in our prior work [5]. After the last stages (**f** or **g** units) of the SC decoders calculate $c$ or $d$, these LLR messages, together with previous path metrics, are input to $L$ metric computation units (MCUs) to generate $2L$ path metric candidates. Then a sorting block is used to select $L$ largest metrics among the $2L$ candidates. The $L$ paths which are associated with the $L$ selected metrics become the updated survival paths.

Besides the above mentioned computation blocks, the LLR-SCL decoder also contains three types of memory banks. The LLR messages memory bank stores and provides the LLR messages which are used in $L$ LLR-SC decoders. Survival path memory bank and path metrics memory bank store and update the $L$ survival paths and the associated path metrics.

Considering the design of memory bank is straightforward, in this section we focus the discussion on **f**/**g** units, MCU and sorting block.

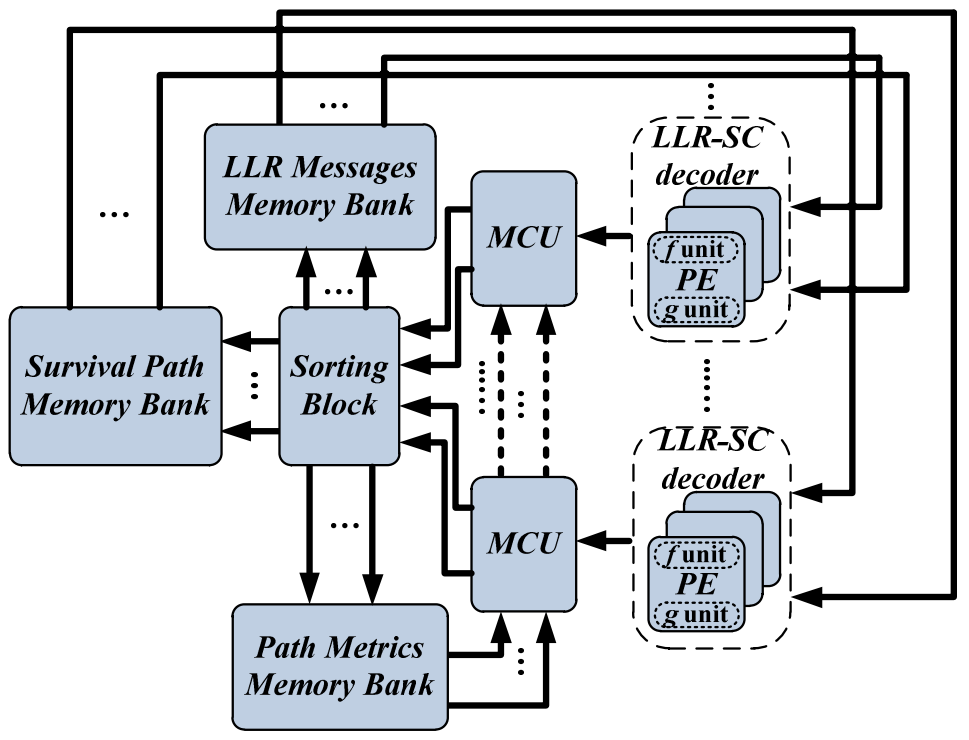

Fig. 6. The overall architecture of LLR-SCL decoder.

### *B. Processing element (PE) for LLR-based f and g units*

As shown in Fig. 6, the LLR-SC component decoder consists of multiple processing elements (PEs). Each PE contains an LLR-based **f** unit and an LLR-based **g** unit. Since the functions of these two units have been described in equations (6)(7), hence the architecture of a $q$-bit PE is developed as shown in Fig. 7. Here C2S and S2C represent the conversion blocks between 2's complement and sign-magnitude forms. The detail of LLR-based PE can be referred to [5].

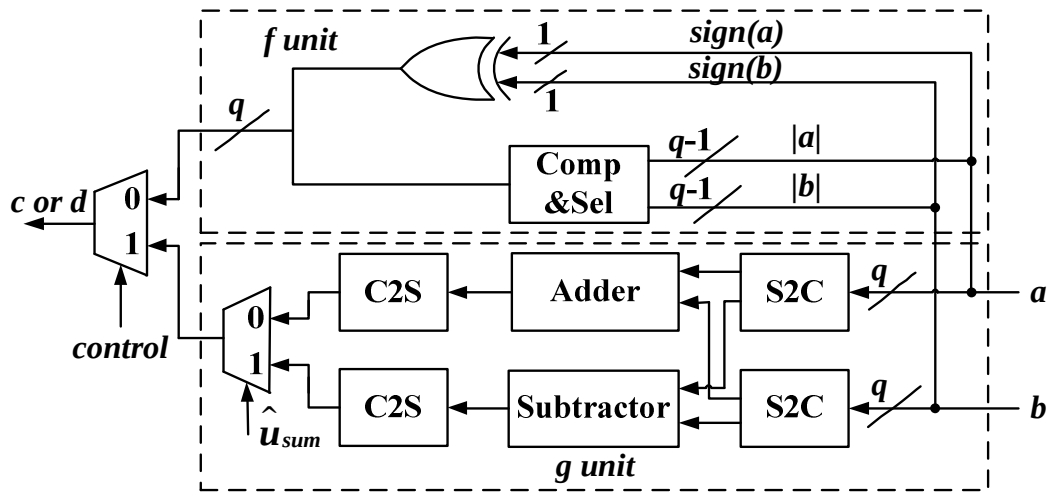

Fig. 7. The architecture of $q$-bit LLR-based PE, cited from [5].

### *C. Metric Computation Unit (MCU)*

(14)-(17) describe the function of MCU. Since for each decoded bit $\hat{u}_i$ either $c$ or $d$ can be input to the MCU, we use *inputLLR* to represent these for convenience. Then, according to (14)-(17), the $q$-bit architecture of MCU is developed as shown in Fig. 8.

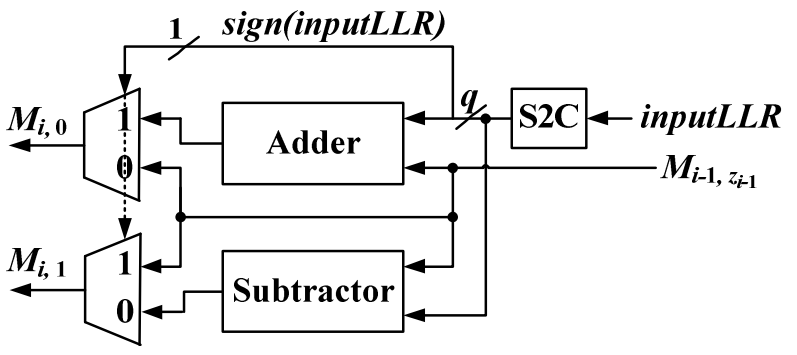

Fig. 8. The architecture of $q$-bit MCU.

### D. Sorting block

Recall that the sorting block is used to compare those metrics and select the $L$ paths with larger metrics. Here we use the batcher odd-even merge algorithm [6] to perform sorting function. Fig. 9 is an example architecture for 8-input sorting. Here *C&S* unit represents the compare and swap operation. It can be seen that for the example 8-input sorting block its critical path delay is 1+2+3=6 $T_{C\&S}$, where $T_{C\&S}$ is the critical path delay of C&S unit. In general, for $2^i$-input sorting block, the critical path delay is $1+2+...i=(i+1)i/2\ T_{C\&S}$, and it is also the critical path delay of the overall LLR-SCL decoder.

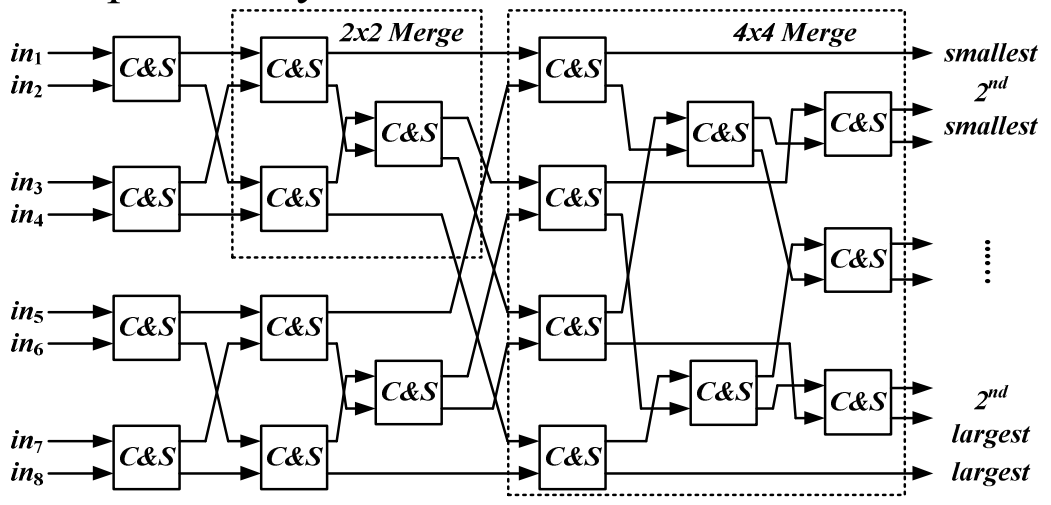

Fig. 9. The architecture of 8-input sorting block.

## V. Analysis of Hardware Performance

In this section the hardware performance of the LLR-SCL decoder is analyzed. Table I lists the hardware performances of LLR-SCL decoder and the conventional log-likelihood-based SCL (LL-SCL) decoder [3-4]. For fair comparison, we assume that the listed decoders have the same $n$=1024, $L$=4 and $q$-bit quantization[1]. In addition, both of their SC component decoders adopt line-type architecture [7].

Table I shows that the proposed LLR-SCL decoder is very area efficient than prior designs. Compared with the LL-SCL architecture, the LLR-SCL decoder reduces total gate count by 50%. This great saving is contributed to the reduced need of data storage and computation for soft messages. Moreover, because the LLR-SCL and LL-SCL decoders have the same critical path delay and latency, it means these two designs have the same throughput. Therefore, hardware efficiency of LLR-SCL decoder, defined as the ratio between throughput and gate count, is 98% higher than that of LL-SCL design.

It is noticed that another approach that derives the LLR-based SCL was reported in [9]. Without prior access to [9], we independently propose the derivation procedure in this paper. Our derivation procedure is different from [9] but leads to the same final LLR-SCL form. This illustrates that the inherent procedure of SCL algorithms can be interpreted in different ways.

[1] Notice in practical design, LLR-based decoder usually needs less quantized bits than LL-based decoder for the same decoding performance. This phenomenon further demonstrates the advantage of the LLR-based solution.

TABLE I. Comparison Between LLR-SCL and LL-SCL Decoders

| Design | | Proposed LLR-SCL | Original LL-SCL |
|---|---|---|---|
| # of PE | | $Ln/2$=2048 | |
| PE | # of Adder | 2 | 4 |
| | # of C2S/S2C | 4 | 8 |
| | # of Comp&Sel | 1 | 2 |
| | # of MUX | 2 | 4 |
| | Gate count | ~17$q$ | ~34$q$ |
| # of MCU | | $L$=4 | 0 |
| MCU | # of Adder | 2 | N/A |
| | # of MUX | 2 | N/A |
| | # of C2S/S2C | 1 | N/A |
| | Gate count | ~25/2$q$ | N/A |
| # of Sorting Block | | 1 | |
| Sorting Block | # of C&S | 19 | |
| | Gate count | ~76$q$ | |
| Bits of LLR/LL memory bank | | ~$L(2n-1)q$ =8188$q$ | ~$L(4n-2)q$ =16376$q$ |
| Bits of Path metrics memory bank | | ~$Lq$=4$q$ | |
| Bits of survival path memory bank | | ~$Ln$=4096 | |
| Total Gate count | | ~43134$q$+4096 | ~86088$q$+4096 |
| Critical path delay | | 6$T_{C\&S}$ | |
| Latency (clock cycle) | | 3$n$-2=3070 | |
| Throughput (Normalized) | | 1 | 1 |
| Hardware Efficiency (Normalized) | | 1.98 | 1 |

## VI. Conclusion

This paper presents LLR-based SCL decoding algorithm and hardware architecture. Analysis results show that the proposed LLR-SCL decoder has very low hardware complexity and is suitable for current LLR-based systems.